\documentclass{article}

\usepackage[numbers]{natbib}
\usepackage[preprint]{neurips_2022}

\usepackage[utf8]{inputenc} 
\usepackage[T1]{fontenc}    
\usepackage{hyperref}       
\usepackage{url}            
\usepackage{booktabs}       
\usepackage{amsfonts}       
\usepackage{nicefrac}       
\usepackage{microtype}      
\usepackage[dvipsnames]{xcolor}
\usepackage{amsmath}
\usepackage{graphicx}
\usepackage{graphics}
\usepackage{subcaption}
\usepackage{accents}
\usepackage{enumitem}
\usepackage{tabu}
\usepackage{wrapfig}
\usepackage{graphicx}
\usepackage{multirow}
\usepackage{multicol}
\usepackage{float}
\usepackage{color}
\usepackage{caption}
\usepackage{bbding}
\usepackage{hanging}

\def\myname{NaturalSpeech 2}

\title{\textit{NaturalSpeech 2}: Latent Diffusion Models are Natural and Zero-Shot Speech and Singing Synthesizers}

\author{
Kai Shen\thanks{The first three authors contributed equally to this work, and their names are listed in random order. Corresponding author: Xu Tan, \texttt{xuta@microsoft.com}}, ~Zeqian Ju\footnotemark[1], ~Xu Tan\footnotemark[1], ~Yanqing Liu, Yichong Leng, Lei He \\
\textbf{Tao Qin, Sheng Zhao, Jiang Bian} \\ 
Microsoft Research Asia \& Microsoft Azure Speech \\
\url{https://aka.ms/speechresearch} 
}
\begin{document}

\maketitle

\vspace{-0.3cm}

\begin{abstract}

Scaling text-to-speech (TTS) to large-scale, multi-speaker, and in-the-wild datasets is important to capture the diversity in human speech such as speaker identities, prosodies, and styles (e.g., singing). Current large TTS systems usually quantize speech into discrete tokens and use language models to generate these tokens one by one, which suffer from unstable prosody, word skipping/repeating issue, and poor voice quality. In this paper, we develop \textit{NaturalSpeech 2}, a TTS system that leverages a neural audio codec with residual vector quantizers to get the quantized latent vectors and uses a diffusion model to generate these latent vectors conditioned on text input. To enhance the zero-shot capability that is important to achieve diverse speech synthesis, we design a speech prompting mechanism to facilitate in-context learning in the diffusion model and the duration/pitch predictor. We scale NaturalSpeech 2 to large-scale datasets with 44K hours of speech and singing data and evaluate its voice quality on unseen speakers. NaturalSpeech 2 outperforms previous TTS systems by a large margin in terms of prosody/timbre similarity, robustness, and voice quality in a zero-shot setting, and performs novel zero-shot singing synthesis with only a speech prompt. Audio samples are available at \url{https://speechresearch.github.io/naturalspeech2}.

\end{abstract}

\begin{figure*}[ht]
  \centering
  \includegraphics[page=1,width=0.95\columnwidth,trim=0cm 4.2cm 7.8cm 0cm,clip=true]{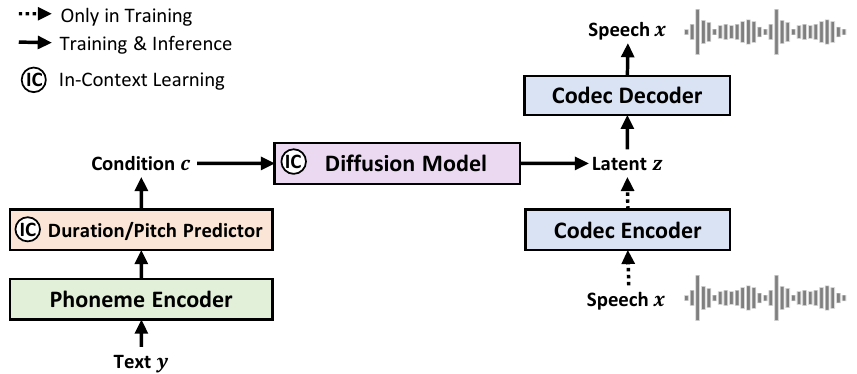}
  \caption{The overview of NaturalSpeech 2, with an audio codec encoder/decoder and a latent diffusion model conditioned on a prior (a phoneme encoder and a duration/pitch predictor). The details of in-context learning in the duration/pitch predictor and diffusion model are shown in Figure~\ref{fig_incontext}.}
  \label{fig_system_overview}
\end{figure*}

\section{Introduction}
\label{sec_intro}

Human speech is full of diversity, with different speaker identities (e.g., gender, accent, timbre), prosodies, styles (e.g., speaking, singing), etc. Text-to-speech (TTS)~\cite{taylor2009text,tan2021survey} aims to synthesize natural and human-like speech with both good quality and diversity. With the development of neural networks and deep learning, TTS systems~\cite{wang2017tacotron,shen2018natural,li2019neural,ren2019fastspeech,ren2021fastspeech,liu2021delightfultts,liu2022delightfultts,kim2021conditional,tan2022naturalspeech} have achieved good voice quality in terms of intelligibility and naturalness, and some systems (e.g., NaturalSpeech~\cite{tan2022naturalspeech}) even achieves human-level voice quality on single-speaker recording-studio benchmarking datasets (e.g., LJSpeech~\cite{ljspeech17}). Given the great achievements in speech intelligibility and naturalness made by the whole TTS community, now we enter a new era of TTS where speech diversity becomes more and more important in order to synthesize natural and human-like speech.

Previous speaker-limited recording-studio datasets are not enough to capture the diverse speaker identities, prosodies, and styles in human speech due to limited data diversity. Instead, we can train TTS models on a large-scale corpus to learn these diversities, and as a by-product, these trained models can generalize to the unlimited unseen scenarios with few-shot or zero-shot technologies. Current large-scale TTS systems~\cite{wang2023neural,kharitonov2023speak,xue2023foundationtts} usually quantize the continuous speech waveform into discrete tokens and model these tokens with autoregressive language models. This pipeline suffers from several limitations: 1) The speech (discrete token) sequence is usually very long (a 10s speech usually has thousands of discrete tokens) and the autoregressive models suffer from error propagation and thus unstable speech outputs. 2) There is a dilemma between the codec and language model: on the one hand, the codec with token quantization (VQ-VAE~\cite{van2017neural,razavi2019generating} or VQ-GAN~\cite{esser2021taming}) usually has a low bitrate token sequence, which, although eases the language model generation, incurs information loss on the high-frequency fine-grained acoustic details; on the other hand, some improving methods~\cite{zeghidour2021soundstream,defossez2022high} use multiple residual discrete tokens to represent a speech frame, which increases the length of the token sequence multiple times if flattened and incurs difficulty in language modeling.

In this paper, we propose \textit{\myname{}}, a TTS system with latent diffusion models to achieve expressive prosody, good robustness, and most importantly strong zero-shot ability for speech synthesis. As shown in Figure~\ref{fig_system_overview}, we first train a neural audio codec that converts a speech waveform into a sequence of latent vectors with a codec encoder, and reconstructs the speech waveform from these latent vectors with a codec decoder. After training the audio codec, we use the codec encoder to extract the latent vectors from the speech in the training set and use them as the target of the latent diffusion model, which is conditioned on prior vectors obtained from a phoneme encoder, a duration predictor, and a pitch predictor. During inference, we first generate the latent vectors from the text/phoneme sequence using the latent diffusion model and then generate the speech waveform from these latent vectors using the codec decoder. 

\begin{table}[h!]
\small
	\centering
    \caption{The comparison between NaturalSpeech 2 and previous large-scale TTS systems.}
	\begin{tabular}{l l l}
	\toprule
     Methods & Previous Systems~\cite{wang2023neural,kharitonov2023speak,xue2023foundationtts} & NaturalSpeech 2 \\
     \midrule
    Representations & Discrete Tokens  & Continuous Vectors \\
    Generative Models & Autoregressive Models & Non-Autoregressvie/Diffusion \\
    In-Context Learning & Both Text and Speech are Needed & Only Speech is Needed \\
    \midrule
    Stability/Robustness? & \XSolidBrush  & \Checkmark \\   
    One Acoustic Model?& \XSolidBrush  & \Checkmark \\  
    Beyond Speech (e.g., Singing)? & \XSolidBrush  & \Checkmark \\  
    \bottomrule
    \end{tabular}
     \label{tab_comparison_tts_system}
\end{table}

We elaborate on some design choices in NaturalSpeech 2 (shown in Table~\ref{tab_comparison_tts_system}) as follows.
\begin{itemize}[leftmargin=*]
    \item \textit{Continuous vectors instead of discrete tokens.} To ensure the speech reconstruction quality of the neural codec, previous works usually quantize speech with multiple residual quantizers. As a result, the obtained discrete token sequence is very long (e.g., if using 8 residual quantizers for each speech frame, the resulting flattened token sequence will be 8 times longer), and puts much pressure on the acoustic model (autoregressive language model). Therefore, we use continuous vectors instead of discrete tokens, which can reduce the sequence length and increase the amount of information for fine-grained speech reconstruction (see Section~\ref{sec_method_codec}).
    
    \item \textit{Diffusion models instead of autoregressive models.} We leverage diffusion models to learn the complex distributions of continuous vectors in a non-autoregressive manner and avoid error propagation in autoregressive models (see Section~\ref{sec_method_diff}). 
    
    \item \textit{Speech prompting mechanisms for in-context learning.} To encourage the diffusion models to follow the characteristics in the speech prompt and enhance the zero-shot capability, we design speech prompting mechanisms to facilitate in-context learning in the diffusion model and pitch/duration predictors (see Section~\ref{sec_method_incontext}). 

\end{itemize}

Benefiting from these designs, NaturalSpeech 2 is more stable and robust than previous autoregressive models, and only needs one acoustic model (the diffusion model) instead of two-stage token prediction as in ~\cite{borsos2022audiolm,wang2023neural}, and can extend the styles beyond speech (e.g., singing voice) due to the duration/pitch prediction and non-autoregressive generation.  

We scale NaturalSpeech 2 to 400M model parameters and 44K hours of speech data, and generate speech with diverse speaker identities, prosody, and styles (e.g., singing) in zero-shot scenarios (given only a few seconds of speech prompt). Experiment results show that NaturalSpeech 2 can generate natural speech in zero-shot scenarios and outperform the previous strong TTS systems. Specifically, 1) it achieves more similar prosody with both the speech prompt and ground-truth speech; 2) it achieves comparable or better naturalness (in terms of CMOS) than the ground-truth speech on LibriSpeech and VCTK test sets; 3) it can generate singing voices in a novel timbre either with a short singing prompt, or interestingly with only a speech prompt, which unlocks the truly zero-shot singing synthesis (without a singing prompt). Audio samples can be found in \url{https://speechresearch.github.io/naturalspeech2}.

\section{Background}

We introduce some background of \myname{}, including the journey of text-to-speech synthesis on pursuing natural voice with high quality and diversity, neural audio codec models, and generative models for audio synthesis. 

\subsection{TTS for Natural Voice: Quality and Diversity}
\label{sec_background_quality_diversity}
Text-to-speech systems~\cite{tan2021survey,wang2017tacotron,shen2018natural,li2019neural,ren2019fastspeech,liu2021delightfultts,liu2022delightfultts,liu2022diffsinger,kim2021conditional,tan2022naturalspeech} aim to generate natural voice with both high quality and diversity. While previous neural TTS systems can synthesize high-quality voice on single-speaker recording-studio datasets (e.g., LJSpeech~\cite{ljspeech17}) and even achieve human-level quality (e.g., NaturalSpeech~\cite{tan2022naturalspeech}), they cannot generate diverse speech with different speaker identities, prosodies, and styles, which are critical to ensure the naturalness of the synthesized speech. Thus, some recent works~\cite{wang2023neural,kharitonov2023speak,xue2023foundationtts} attempt to scale the TTS systems to large-scale, multi-speaker, and in-the-wild datasets to pursue diversity.

These systems usually leverage a neural codec to convert speech waveform into discrete token sequence and an autoregressive language model to generate discrete tokens from text, which suffers from a dilemma as shown in Table~\ref{tab_delimma}: 1) If the audio codec quantizes each speech frame into a single token with vector-quantizer (VQ)~\cite{van2017neural,razavi2019generating,esser2021taming}, this could ease the token generation in the language model due to short sequence length, but will affect the waveform reconstruction quality due to large compression rate or low bitrate. 2) If the audio codec quantizes each speech frame into multiple tokens with residual vector-quantizer (RVQ)~\cite{zeghidour2021soundstream,defossez2022high}, this will ensure high-fidelity waveform reconstruction, but will cause difficulty in autoregressive model generation (error propagation and robust issues) due to the increased length in the token sequence. Thus, previous works such as AudioLM~\cite{borsos2022audiolm} leverage two-stage language models to first generate some coarse-grained tokens in each frame and then generate the remaining fine-grained tokens, which are complicated and incur cascaded errors. To avoid the above dilemma, we leverage a neural codec with continuous vectors and a latent diffusion model with non-autoregressive generation.

\begin{table}[h!]
\small
\caption{The dilemma in the pipeline of discrete audio codec and autoregressive language model.}
\centering
\begin{tabular}{l |c c}
\toprule
The Dilemma in Previous Systems & Single Token (VQ) & Multiple Tokens (RVQ) \\
\midrule
Waveform Reconstruction (Discrete Audio Codec) & Hard & Easy  \\
Token Generation (Autoregressive Language Model) & Easy & Hard \\
\bottomrule
\end{tabular}
\label{tab_delimma}
\end{table}

\subsection{Neural Audio Codec}
\label{sec_background_codec}

Neural audio codec~\cite{oord2016wavenet,valin2019lpcnet,zeghidour2021soundstream,defossez2022high} refers to a kind of neural network model that converts audio waveform into compact representations with a codec encoder and reconstructs audio waveform from these representations with a codec decoder. Since audio codec is traditionally used for audio compression and transmission, the compression rate is a critical metric and thus discrete tokens with low bitrate are usually chosen as the compact representations. For example, SoundStream~\cite{zeghidour2021soundstream} and Encodec~\cite{defossez2022high} leverage vector-quantized variational auto-encoders (VQ-VAE) with multiple residual vector-quantizers to compress speech into multiple tokens, and have been used as the intermediate representations for speech/audio generation~\cite{borsos2022audiolm,kreuk2022audiogen,wang2023neural,kharitonov2023speak,xue2023foundationtts}. 

Although good reconstruction quality and low bitrate can be achieved by residual vector quantizers, they are mainly designed for compression and transmission purposes and may not be suitable to serve as the intermediate representation for speech/audio generation. The discrete token sequence generated by residual quantizers is usually very long ($R$ times longer if $R$ residual quantizers are used), which is difficult for the language models to predict. Inaccurate predictions of discrete tokens will cause word skipping, word repeating, or speech collapse issues when reconstructing speech waveforms from these tokens. In this paper, we design a neural audio codec to convert speech waveform into continuous vectors instead of discrete tokens, which can maintain enough fine-grained details for precise waveform reconstruction without increasing the length of the sequence.

\subsection{Generative Models for Speech Synthesis}
\label{sec_background_dgm}

Different generative models have been applied to speech or audio synthesis, and among these, autoregressive models and diffusion models are the two most prominent methods. Autoregressive models have long been used in speech synthesis for waveform generation~\cite{oord2016wavenet} or acoustic feature generation~\cite{wang2017tacotron}. Inspired by the success of autoregressive models in language generation~\cite{radford2018improving,radford2019language,brown2020language}, autoregressive models have been applied in speech and audio generation~\cite{borsos2022audiolm,kreuk2022audiogen,wang2023neural,kharitonov2023speak,xue2023foundationtts}. Meanwhile, diffusion models have also been widely used in speech synthesis for waveform generation~\cite{kong2020diffwave,chen2020wavegrad} and acoustic feature generation~\cite{jeong2021diff,popov2021grad}. 

Although both models are based on iterative computation (following the left-to-right process or the denoising process), autoregressive models are more sensitive to sequence length and error propagation, which cause unstable prosody and robustness issues (e.g., word skipping, repeating, and collapse). Considering text-to-speech has a strict monotonic alignment and strong source-target dependency, we leverage diffusion models enhanced with duration prediction and length expansion, which are free from robust issues.

\section{NaturalSpeech 2}
\label{sec_method}

In this section, we introduce NaturalSpeech 2, a TTS system for natural and zero-shot voice synthesis with high fidelity/expressiveness/robustness on diverse scenarios (various speaker identities, prosodies, and styles). As shown in Figure~\ref{fig_system_overview}, NaturalSpeech 2 consists of a neural audio codec (an encoder and a decoder) and a diffusion model with a prior (a phoneme encoder and a duration/pitch predictor). Since speech waveform is complex and high-dimensional, following the paradigm of regeneration learning~\cite{tan2023regeneration}, we first convert speech waveform into latent vectors using the audio codec encoder and reconstruct speech waveform from the latent vectors using the audio codec decoder. Next, we use a diffusion model to predict the latent vectors conditioned on text/phoneme input. We introduce the detailed designs of neural audio codec in Section~\ref{sec_method_codec} and the latent diffusion model in Section~\ref{sec_method_diff}, as well as the speech prompting mechanism for in-context learning in Section~\ref{sec_method_incontext}.

\subsection{Neural Audio Codec with Continuous Vectors}
\label{sec_method_codec}

We use a neural audio codec to convert speech waveform into continuous vectors instead of discrete tokens, as analyzed in Section~\ref{sec_background_quality_diversity} and~\ref{sec_background_codec}. Audio codec with continuous vectors enjoys several benefits: 1) Continuous vectors have a lower compression rate and higher bitrate than discrete tokens\footnote{Since our task is not speech compression but speech synthesis, we do not need a high compression rate or a low bitrate.}, which can ensure high-quality audio reconstruction. 2) Each audio frame only has one vector instead of multiple tokens as in discrete quantization, which will not increase the length of the hidden sequence.

\begin{figure*}[t]
  \centering
  \includegraphics[page=2,width=1\columnwidth,trim=0cm 5.2cm 4.5cm 0cm,clip=true]{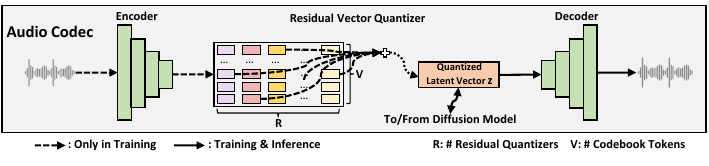}
  \caption{The neural audio codec consists of an encoder, a residual vector-quantizer (RVQ), and a decoder. The encoder extracts the frame-level speech representations from the audio waveform, the RVQ leverages multiple codebooks to quantize the frame-level representations, and the decoder takes the quantized vectors as input and reconstructs the audio waveform. The quantized vectors also serve as the training target of the latent diffusion model.}
  \label{fig_codec}
\end{figure*}

As shown in Figure~\ref{fig_codec}, our neural audio codec consists of an audio encoder, a residual vector-quantizer (RVQ), and an audio decoder: 1) The audio encoder consists of several convolutional blocks with a total downsampling rate of $200$ for $16$KHz audio, i.e., each frame corresponds to a $12.5$ms speech segment. 2) The residual vector-quantizer converts the output of the audio encoder into multiple residual vectors following~\cite{zeghidour2021soundstream}. The sum of these residual vectors is taken as the quantized vectors, which are used as the training target of the diffusion model. 3) The audio decoder mirrors the structure of the audio encoder, which generates the audio waveform from the quantized vectors. 
The working flow of the neural audio codec is as follows.
\begin{equation} 
\begin{aligned}
&{\rm Audio~Encoder}:  h = f_{\rm enc} (x),\\
&{\rm Residual~Vector~Quantizer}: \{e^i_j\}^R_{j=1} = f_{\rm rvq} (h^i), ~~z^i = \sum\limits^R_{j=1}e^i_j, ~~z = \{z^i\}^n_{i=1},  \\
&{\rm Audio~Decoder}: x  = f_{\rm dec} (z),\\
\label{eq_codec}
\end{aligned}
\end{equation}
where $f_{\rm enc}$, $f_{\rm rvq}$, and $f_{\rm dec}$ denote the audio encoder, residual vector quantizer, and audio decoder. $x$ is the speech waveform, $h$ is the hidden sequence obtained by the audio encoder with a frame length of $n$, and $z$ is the quantized vector sequence with the same length as $h$. $i$ is the index of the speech frame, $j$ is the index of the residual quantizer and $R$ is the total number of residual quantizers, and $e^i_j$ is the embedding vector of the codebook ID obtained by the $j$-th residual quantizer on the $i$-th hidden frame (i.e.,  $h^i$). The training of the neural codec follows the loss function in~\cite{zeghidour2021soundstream}.

Actually, to obtain continuous vectors, we do not need vector quantizers, but just an autoencoder or variational autoencoder. However, for regularization and efficiency purposes, we use residual vector quantizers with a very large number of quantizers ($R$ in Figure~\ref{fig_codec}) and codebook tokens ($V$ in Figure~\ref{fig_codec}) to approximate the continuous vectors. By doing this, we have two benefits: 1) When training latent diffusion models, we do not need to store continuous vectors which are memory-cost. Instead, we just store the codebook embeddings and the quantized token IDs, which are used to derive the continuous vectors using Equation~\ref{eq_codec}. 2) When predicting the continuous vectors, we can add an additional regularization loss on discrete classification based on these quantized token IDs (see $\mathcal{L}_{\rm ce-rvq}$ in Section~\ref{sec_method_diff}).

\subsection{Latent Diffusion Model with Non-Autoregressive Generation}
\label{sec_method_diff}

We leverage a diffusion model to predict the quantized latent vector $z$ conditioned on the text sequence $y$. We leverage a prior model that consists of a phoneme encoder, a duration predictor, and a pitch predictor to process the text input and provide a more informative hidden vector $c$ as the condition of the diffusion model. 

\paragraph{Diffusion Formulation}
We formulate the diffusion (forward) process and denoising (reverse) process as a stochastic differential equation (SDE)~\cite{song2020score}, respectively. The forward SDE transforms the latent vectors $z_0$ obtained by the neural codec (i.e., $z$) into Gaussian noises:
\begin{equation} 
\mathrm{d} z_t = -\frac{1}{2} \beta_t z_t~\mathrm{d}t + \sqrt{\beta_t}~\mathrm{d}w_t, \quad t \in [0,1],
\end{equation}
where $w_t$ is the standard Brownian motion, $t \in [0,1]$, and $\beta_t$ is a non-negative noise schedule function. Then the solution is given by:
\begin{equation}
z_t = e^{-\frac{1}{2}\int_{0}^{t} \beta_s ds}z_0 + \int_{0}^{t} \sqrt{\beta_s}e^{-\frac{1}{2}\int_{0}^{t} \beta_u du} \mathrm{d}w_s.
\end{equation}
By properties of Ito’s integral, the conditional distribution of $z_t$ given $z_0$ is Gaussian: $p(z_t | z_0) \sim \mathcal{N}(\rho(z_0, t), \Sigma_t)$, where $\rho(z_0, t) = e^{-\frac{1}{2}\int_{0}^{t} \beta_s ds}z_0$ and $\Sigma_t = I-e^{-\int_{0}^{t} \beta_s ds}$.

The reverse SDE transforms the Gaussian noise back to data $z_0$ with the following process:
\begin{equation} 
\mathrm{d} z_t = - (\frac{1}{2} z_t + \nabla \log p_t(z_t)) \beta_t ~\mathrm{d}t + \sqrt{\beta_t}~\mathrm{d}\Tilde{w}_t, \quad t \in [0,1],
\label{eq_reverse_sde}
\end{equation}
where $\Tilde{w}$ is the reverse-time Brownian motion. Moreover, we can consider an ordinary differential equation (ODE)~\cite{song2020score} in the reverse process:
\begin{equation} 
\mathrm{d} z_t = -\frac{1}{2} (z_t + \nabla \log p_t(z_t)) \beta_t ~\mathrm{d}t, \quad t \in [0,1].
\label{eq_reverse_ode}
\end{equation}
We can train a neural network $s_{\theta}$ to estimate the score $\nabla \log p_t(z_t)$ (the gradient of the log-density of noisy data), and then we can sample data $z_0$ by starting from Gaussian noise $z_1 \sim \mathcal{N}(0, 1)$ and numerically solving the SDE in Equation~\ref{eq_reverse_sde} or ODE in Equation~\ref{eq_reverse_ode}. In our formulation, the neural network $s_{\theta}(z_t, t, c)$ is based on WaveNet~\cite{oord2016wavenet}, which takes the current noisy vector $z_t$, the time step $t$, and the condition information $c$ as input, and predicts the data $\hat z_0$ instead of the score, which we found results in better speech quality. Thus, $\hat z_0 = s_{\theta}(z_t, t, c)$. The loss function to train the diffusion model is as follows.
\begin{equation} 
\begin{aligned}
\mathcal{L}_{\rm diff} &= \mathbb{E}_{z_0, t}[||\hat z_0 - z_0||_2^2 + ||\Sigma_t^{-1}(\rho(\hat z_0, t) -z_t)- \nabla \log p_t(z_t)||_2^2 + \lambda_{ce-rvq}\mathcal{L}_{\rm ce-rvq}], 
\label{eq_diff_loss}
\end{aligned}
\end{equation}
where the first term is the data loss, the second term is the score loss, and the predicted score is calculated by $\Sigma_t^{-1}(\rho(\hat z_0, t)-z_t)$, which is also used for reverse sampling based on Equation~\ref{eq_reverse_sde} or~\ref{eq_reverse_ode} in inference. The third term $\mathcal{L}_{\rm ce-rvq}$ is a novel cross-entropy (CE) loss based on residual vector-quantizer (RVQ). Specifically, for each residual quantizer $j \in [1, R]$, we first get the residual vector $\hat z_0 - \sum^{j-1}_{i=1}e_i$, where $e_i$ is the ground-truth quantized embedding in the $i$-th residual quantizer ($e_i$ is also introduced in Equation~\ref{eq_codec}). Then we calculate the L2 distance between the residual vector with each codebook embedding in quantizer $j$ and get a probability distribution with a softmax function, and then calculate the cross-entropy loss between the ID of the ground-truth quantized embedding $e_j$ and this probability distribution. $\mathcal{L}_{\rm ce-rvq}$ is the mean of the cross-entropy loss in all $R$ residual quantizers, and $\lambda_{\rm ce-rvq}$ is set to 0.1 during training.

\paragraph{Prior Model: Phoneme Encoder and Duration/Pitch Predictor} The phoneme encoder consists of several Transformer blocks~\cite{vaswani2017attention,ren2019fastspeech}, where the standard feed-forward network is modified as a convolutional network to capture the local dependency in phoneme sequence. Both the duration and pitch predictors share the same model structure with several convolutional blocks but with different model parameters. The ground-truth duration and pitch information is used as the learning target to train the duration and pitch predictors, with an L1 duration loss $\mathcal{L}_{\rm dur}$ and pitch loss $\mathcal{L}_{\rm pitch}$. During training, the ground-truth duration is used to expand the hidden sequence from the phoneme encoder to obtain the frame-level hidden sequence, and then the ground-truth pitch information is added to the frame-level hidden sequence to get the final condition information $c$. During inference, the corresponding predicted duration and pitch are used. 

The total loss function for the diffusion model is as follows:
\begin{equation} 
\mathcal{L} = \mathcal{L}_{\rm diff} + \mathcal{L}_{\rm dur} + \mathcal{L}_{\rm pitch}.
\label{eq_loss}
\end{equation}

\begin{figure*}[ht]
  \centering
  \includegraphics[page=3,width=0.9\columnwidth,trim=0cm 5.4cm 7.9cm 0cm,clip=true]{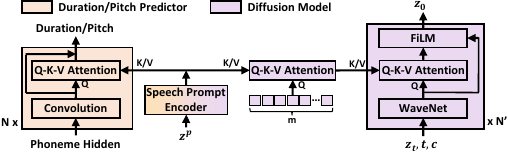}
  \caption{The speech prompting mechanism in the duration/pitch predictor and the diffusion model for in-context learning. During training, we use a random segment $z^{u:v}$ of the target speech $z$ as the speech prompt $z^p$ and use the diffusion model to only predict $z^{\setminus u:v}$. During inference, we use a reference speech of a specific speaker as the speech prompt $z^p$. Note that the prompt is the speech latent obtained by the codec encoder instead of the speech waveform.}
  \label{fig_incontext}
\end{figure*}

\subsection{Speech Prompting for In-Context Learning}
\label{sec_method_incontext}

To facilitate in-context learning for better zero-shot generation, we design a speech prompting mechanism to encourage the duration/pitch predictor and the diffusion model to follow the diverse information (e.g., speaker identities) in the speech prompt. For a speech latent sequence $z$, we randomly cut off a segment $z^{u:v}$ with frame index from $u$ to $v$ as the speech prompt, and concatenate the remaining speech segments $z^{1:u}$ and $z^{v:n}$ to form a new sequence $z^{\setminus u:v}$ as the learning target of the diffusion model. As shown in Figure~\ref{fig_incontext}, we use a Transformer-based prompt encoder to process the speech prompt $z^{u:v}$ ($z^p$ in the figure) to get a hidden sequence. To leverage this hidden sequence as the prompt, we have two different strategies for the duration/pitch predictor and the diffusion model: 1) For the duration and pitch predictors, we insert a Q-K-V attention layer in the convolution layer, where the query is the hidden sequence of the convolution layer, and the key and value is the hidden sequence from the prompt encoder. 2) For the diffusion model, instead of directly attending to the hidden sequence from the prompt encoder that exposes too many details to the diffusion model and may harm the generation, we design two attention blocks: in the first attention block, we use $m$ randomly initialized embeddings as the query sequence to attend to the prompt hidden sequence, and get a hidden sequence with a length of $m$ as the attention results~\cite{wang2016attention,wang2018style,yin2022retrievertts}; in the second attention block, we leverage the hidden sequence in the WaveNet layer as the query and the $m$-length attention results as the key and value. We use the attention results of the second attention block as the conditional information of a FiLM layer~\cite{perez2018film} to perform affine transform on the hidden sequence of the WaveNet in the diffusion model. Please refer to Appendix~\ref{append_wavenet_architecture} for the details of WaveNet architecture used in the diffusion model.

\subsection{Connection to NaturalSpeech}
NaturalSpeech 2 is an advanced edition of the NaturalSpeech Series~\cite{tan2022naturalspeech,shen2023naturalspeech}. Compared to its previous version NaturalSpeech~\cite{tan2022naturalspeech}, NaturalSpeech 2 has the following connections and distinctions. First, \textit{goal}. While both NaturalSpeech 1 and 2 aim at synthesizing natural voices (with good speech quality and diversity), their focuses are different. NaturalSpeech focuses on speech quality by synthesizing voices that are on par with human recordings and only tackling single-speaker recording-studio datasets (e.g., LJSpeech). NaturalSpeech 2 focuses on speech diversity by exploring the zero-shot synthesis ability based on large-scale, multi-speaker, and in-the-wild datasets. Second, \textit{architecture}. NaturalSpeech 2 keeps the basic components in NaturalSpeech, such as the encoder and decoder for waveform reconstruction, and the prior module (phoneme encoder, duration/pitch predictor). However, it leverages 1) a diffusion model to increase the modeling power to capture the complicated and diverse data distribution in large-scale speech datasets, 2) a residual vector quantizer to regularize the latent vectors to trade off the reconstruction quality and prediction difficulty, and 3) a speech prompting mechanism to enable zero-shot ability that is not covered in single-speaker synthesis system.

\section{Experimental Settings}
In this section, we introduce the experimental settings to train and evaluate NaturalSpeech 2, including the dataset, model configuration, baselines for comparison, training and inference, and evaluation metrics. 
\subsection{Datasets}
\label{sec_exp_dataset}
\paragraph{Training Dataset} To train the neural audio codec and the diffusion model, we use the English subset of Multilingual LibriSpeech (MLS)~\cite{pratap2020mls} as the training data, which contains $44$K hours of transcribed speech data derived from LibriVox audiobooks. The number of distinct speakers is $2742$ for males and $2748$ for females respectively. The sample rate is $16$KHz for all speech data. The input text sequence is first converted into a phoneme sequence using grapheme-to-phoneme conversion~\cite{sun2019token} and then aligned with speech using our internal alignment tool to obtain the phoneme-level duration. The frame-level pitch sequence is extracted from the speech using PyWorld\footnote{\url{https://github.com/JeremyCCHsu/Python-Wrapper-for-World-Vocoder}}.

\paragraph{Evaluation Dataset} We employ two benchmark datasets for evaluation: 1) LibriSpeech~\cite{panayotov2015librispeech} test-clean, which contains $40$ distinct speakers and $5.4$ hours of annotated speech data. 2) VCTK dataset~\cite{veaux2016superseded}, which contains $108$ distinct speakers. For LibriSpeech test-clean, we randomly sample $15$ utterances for each speaker and form a subset of 600 utterances for evaluation. For VCTK, we randomly sample $5$ utterances for each speaker, resulting in a subset of $540$ utterances for evaluation. Specifically, to synthesize each sample, we randomly select a different utterance of the same speaker and crop it into a $\sigma$-second audio segment to form a $\sigma$-second prompt.
Note that both the speakers in LibriSpeech test-clean and VCTK are not seen during training. Thus, we aim to conduct zero-shot speech synthesis.

The singing datasets follow a similar process in the speech dataset, and the details are shown in Section~\ref{sec_results_singing}.

\subsection{Model Configuration and Comparison}
\paragraph{Model Configuration}
The phoneme encoder is a 6-layer Transformer~\cite{vaswani2017attention} with 8 attention heads, 512 embedding dimensions, 2048 1D convolution filter size, 9 convolution 1D kernel size, and 0.1 dropout rate. The pitch and duration predictor share the same architecture of 30-layer 1D convolution with ReLU activation and layer normalization, 10 Q-K-V attention layers for in-context learning, which have 512 hidden dimensions and 8 attention heads and are placed every 3 1D convolution layers. We set the dropout to 0.5 in both duration and pitch predictors. For the speech prompt encoder, we use a 6-layer Transformer with 512 hidden size, which has the same architecture as the phoneme encoder. As for the $m$ query tokens in the first Q-K-V attention in the prompting mechanism in the diffusion model (as shown in Figure~\ref{fig_incontext}), we set the token number $m$ to 32 and the hidden dimension to 512.

The diffusion model contains 40 WaveNet layers~\cite{oord2016wavenet}, which consist of 1D dilated convolution layers with 3 kernel size, 1024 filter size, and 2 dilation size. Specifically, we use a FiLM layer~\cite{perez2018film} at every 3 WaveNet layers to fuse the condition information processed by the second Q-K-V attention in the prompting mechanism in the diffusion model. The hidden size in WaveNet is 512, and the dropout rate is 0.2.

More details of the model configurations are shown in Appendix~\ref{append_model_details}.

\paragraph{Model Comparison} We choose the previous zero-shot TTS model YourTTS~\cite{casanova2022yourtts} as the baseline, with the official code and pre-trained checkpoint\footnote{\url{https://github.com/Edresson/YourTTS}}, which is trained on VCTK~\cite{veaux2016superseded}, LibriTTS~\cite{zen2019libritts} and TTS-Portuguese~\cite{casanova2022tts}. We also choose VALL-E~\cite{wang2023neural} that is based on discrete audio codec and autoregressive language model for comparison, which can help demonstrate the advantages of the designs in NaturalSpeech 2. We directly collect some audio samples from its demo page for comparison.

\subsection{Model Training and Inference}
\label{sec_config}

We first train the audio codec using $8$ NVIDIA TESLA V100 16GB GPUs with a batch size of 200 audios per GPU for $440$K steps. We follow the implementation and experimental setting of SoundStream~\cite{zeghidour2021soundstream} and adopt Adam optimizer with $2e-4$ learning rate. Then we use the trained codec to extract the quantized latent vectors for each audio  to train the diffusion model in NaturalSpeech 2.

The diffusion model in NaturalSpeech 2 is trained using $16$ NVIDIA TESLA V100 32GB GPUs with a batch size of $6$K frames of latent vectors per GPU for $300$K steps (our model is still underfitting and longer training will result in better performance). We optimize the models with the AdamW optimizer with $5e-4$ learning rate, $32$k warmup steps following the inverse square root learning schedule.

During inference, for the diffusion model, we find it beneficial to use a temperature $\tau$ and sample the terminal condition $z_T$ from $\mathcal{N}(0, \tau ^{-1} I)$~\cite{popov2021grad}. We set $\tau$ to $1.2^2$. To balance the generation quality and latency, we adopt the Euler ODE solver and set the diffusion steps to $150$.

\subsection{Evaluation Metrics} 
\label{sec_exp_metric}
We use both objective and subjective metrics to evaluate the zero-shot synthesis ability of NaturalSpeech 2 and compare it with baselines.
\paragraph{Objective Metrics} We evaluate the TTS systems with the following objective metrics:
\begin{itemize}[leftmargin=*]
    \item \emph{Prosody Similarity with Prompt.} We evaluate the prosody similarity (in terms of pitch and duration) between the generated speech and the prompt speech, which measures how well the TTS model follows the prosody in speech prompt in zero-shot synthesis. We calculate the prosody similarity with the following steps: 1) we extract phoneme-level duration and pitch from the prompt and the synthesized speech; 2) we calculate the mean, standard variation, skewness, and kurtosis~\cite{ren2021fastspeech} of the pitch and duration in each speech sequence; 3) we calculate the difference of the mean, standard variation, skewness, and kurtosis between each paired prompt and synthesized speech and average the differences among the whole test set. 
    \item \emph{Prosody Similarity with Ground Truth.} We evaluate the prosody similarity (in terms of pitch and duration) between the generated speech and the ground-truth speech, which measures how well the TTS model matches the prosody in the ground truth. Since there is correspondence between two speech sequences, we calculate the Pearson correlation and RMSE of the pitch/duration between the generated and ground-truth speech, and average them on the whole test set. 
    \item \emph{Word Error Rate.} We employ an ASR model to transcribe the generated speech and calculate the word error rate (WER). The ASR model is a CTC-based HuBERT~\cite{hsu2021hubert} pre-trained on Librilight~\cite{kahn2020libri} and fine-tuned on the 960 hours training set of LibriSpeech. We use the official code and checkpoint\footnote{\url{https://huggingface.co/facebook/hubert-large-ls960-ft}}. 
\end{itemize}

\paragraph{Subjective Metrics}
We conduct human evaluation and use the intelligibility score and mean opinion score as the subjective metrics:
\begin{itemize}[leftmargin=*]
  \item \emph{Intelligibility Score.} Neural TTS models often suffer from the robustness issues such as word skipping, repeating, and collapse issues, especially for autoregressive models. To demonstrate the robustness of NaturalSpeech 2, following the practice in~\cite{ren2019fastspeech}, we use the 50 particularly hard sentences (see Appendix~\ref{appendix_50hard}) and conduct an intelligibility test. We measure the number of repeating words, skipping words, and error sentences as the intelligibility score. 
  \item \emph{CMOS and SMOS.} Since synthesizing natural voices is one of the main goals of NaturalSpeech 2, we measure naturalness using comparative mean option score (CMOS) with 12 native speakers as the judges. We also use the similarity mean option score (SMOS) between the synthesized and prompt speech to measure the speaker similarity, with 6 native speakers as the judges.
\end{itemize}

\section{Results on Natural and Zero-Shot Synthesis}
In this section, we conduct a series of experiments to compare the \myname{} with the baselines from the following aspects: 1) \textit{Generation Quality}, by evaluating the naturalness of the synthesized audio; 2) \textit{Generation Similarity}, by evaluating how well the TTS system follows prompts; 3) \textit{Robustness}, by calculating the WER and an additional intelligibility test.

\subsection{Generation Quality} 

\begin{wraptable}{r}{0.43\textwidth} \small
\centering
\vspace{-10px}
\caption{The CMOS results (v.s. NaturalSpeech 2) on LibriSpeech and VCTK.}
    \begin{tabular}{lcc}
    \toprule
    Setting & LibriSpeech & VCTK \\
    \midrule
      Ground Truth                         & $+0.04$   & $-0.30$ \\
      \midrule
      YourTTS   & $-0.65$   & $-0.58$ \\
      NaturalSpeech 2                      & $~~~\textbf{0.00}$ & $~~~\textbf{0.00}$ \\
      \bottomrule
    \end{tabular}
\vspace{-10px}
\label{tab_cmos}
\end{wraptable}

We conduct CMOS test to evaluate the generation quality (i.e., naturalness). We randomly select 20 utterances from the LibriSpeech and VCTK tests and crop the prompt speech to 3s. To ensure high-quality generation, we use a speech scoring model~\cite{chen2022wavlm} to filter the multiple samples generated by the diffusion model with different starting Gaussian noises $z_1$. Table \ref{tab_cmos} shows a comparison of NaturalSpeech 2 against baseline YourTTS and the ground truth. We have several observations: 1) NaturalSpeech 2 is comparable to the ground-truth recording in LibriSpeech ($+0.04$ is regarded as on par) and achieves much better quality on VCTK datasets ($-0.30$ is a large gap), which demonstrates the naturalness of the speech generated by NaturalSpeech 2 is high enough. 2) NaturalSpeech shows $0.65$ and $0.58$ CMOS gain over YourTTS in LibriSpeech and VCTK, respectively, which shows the superiority of NaturalSpeech 2 over this baseline.

\subsection{Generation Similarity} 
\label{sec:generation_similarity_exp}

\begin{table}[ht]
\small
\centering
\caption{The prosody similarity between  synthesized and prompt speech in terms of the difference in  mean (Mean), standard variation (Std), skewness (Skew), and kurtosis (Kurt) of pitch and duration.}
\begin{tabular}{lcccccccc}
\toprule
\multirow{2}{*}{\textbf{LibriSpeech}}& \multicolumn{4}{c}{Pitch} & \multicolumn{4}{c}{Duration} \\
\cmidrule{2-5} \cmidrule{6-9}
& Mean$\downarrow$ & Std$\downarrow$ & Skew$\downarrow$ & Kurt$\downarrow$ &  Mean$\downarrow$ & Std$\downarrow$ & Skew$\downarrow$ & Kurt$\downarrow$ \\
\midrule
YourTTS 
 & 10.52 & 7.62 & 0.59 & 1.18 & 0.84 & \textbf{0.66} & 0.75 & 3.70 \\
NaturalSpeech 2 & \textbf{10.11} & \textbf{6.18} & \textbf{0.50} & \textbf{1.01} & \textbf{0.65} & 0.70 & \textbf{0.60} & \textbf{2.99}\\
\midrule
\midrule
\multirow{2}{*}{\textbf{VCTK}} & \multicolumn{4}{c}{Pitch} & \multicolumn{4}{c}{Duration} \\
\cmidrule{2-5} \cmidrule{6-9}
& Mean$\downarrow$ & Std$\downarrow$ & Skew$\downarrow$ & Kurt$\downarrow$ &  Mean$\downarrow$ & Std$\downarrow$ & Skew$\downarrow$ & Kurt$\downarrow$ \\
\midrule
YourTTS
 & 13.67 & 6.63 & 0.72 & 1.54 & \textbf{0.72} & 0.85 & 0.84 & 3.31\\
NaturalSpeech 2 & \textbf{13.29} & \textbf{6.41} & \textbf{0.68} & \textbf{1.27} & 0.79 & \textbf{0.76} & \textbf{0.76} & \textbf{2.65} \\
\bottomrule
\end{tabular}
\label{table:prosody_similarity_syn_prompt}
\end{table}

We use two metrics to evaluate the speech similarity: 1) prosody similarity between the synthesized and prompt speech. 2) SMOS test. To evaluate the prosody similarity, we randomly sample one sentence for each speaker for both LibriSpeech test-clean and VCTK dataset to form the test sets. Specifically, to synthesize each sample, we randomly and independently sample the prompt speech with $\sigma = 3$ seconds. Note that YourTTS has seen $97$ speakers in VCTK in training, but we still compare NaturalSpeech 2 with YourTTS on all the speakers in VCTK (i.e., the 97 speakers are seen to YourTTS but unseen to NaturalSpeech 2). 

\begin{wraptable}{r}{0.43\textwidth} \small
    \centering
    \vspace{-10px}
    \caption{The SMOS on LibriSpeech and VCTK respectively.}
    \begin{tabular}{lcc}
    \toprule
    Setting & LibriSpeech & VCTK  \\
    \midrule
      GroundTruth     &  $3.33$ &  $3.86$  \\
      \midrule
      YourTTS         & $2.03$  &  $2.43$  \\
      NaturalSpeech 2 & $\textbf{3.28}$  &  $\textbf{3.20}$  \\
      \bottomrule
    \end{tabular}
    \label{tab_smos}
    \vspace{-10px}
\end{wraptable}
We apply the alignment tool to obtain phoneme-level duration and pitch and calculate the prosody similarity metrics between the synthesized speech and the prompt speech as described in Section~\ref{sec_exp_metric}. The results are shown in Table \ref{table:prosody_similarity_syn_prompt}. We have the following observations: 1) NaturalSpeech 2 consistently outperforms the baseline YourTTS in both LibriSpeech and VCTK on all metrics, which demonstrates that our proposed NaturalSpeech 2 can mimic the prosody of prompt speech much better. 2) Although YourTTS has seen 97 from 108 speakers in VCTK dataset, our model can still outperform it by a large margin, which demonstrates the advantages of NaturalSpeech 2. Furthermore, we also compare prosody similarity between synthesized and ground-truth speech in Appendix \ref{app:prosody_similarity_with_gt}.

We further evaluate the speaker similarity using SMOS test. We randomly select 10 utterances from LibriSpeech and VCTK datasets respectively, following the setting in the CMOS test. The length
of the prompt speech is set to 3s. The results are shown in Table \ref{tab_smos}. NaturalSpeech 2 outperforms YourTTS by $1.25$ and $0.77$ SMOS scores for LibriSpeech and VCTK, respectively, which shows that NaturalSpeech 2 is significantly better in speaker similarity.

\begin{wraptable}{r}{0.43\textwidth} \small
\centering
\vspace{-10px}
\caption{Word error rate on LibriSpeech and VCTK.}
\begin{tabular}{lcc}
\toprule
Setting & LibriSpeech  & VCTK \\
\midrule
Ground Truth & 1.94 & 9.49\\
\midrule
YourTTS &  7.10 & 14.80\\
NaturalSpeech 2 &  \textbf{2.26} & \textbf{6.99} \\
\bottomrule
\end{tabular}
\label{tab_wer}
\vspace{-10px}
\end{wraptable}

\subsection{Robustness} 
\label{sec:robustness}
We use the full test set of LibriSpeech and VCTK as described in Section~\ref{sec_exp_dataset} to synthesize the speech and compute the word error rate (WER) between the transcribed text and ground-truth text. To synthesize each sample, we use a 3-second prompt by randomly cropping the whole prompt speech. The results are shown in Table \ref{tab_wer}. We observe that: 1) NaturalSpeech 2 significantly outperforms YourTTS in LibriSpeech and VCTK, indicating better synthesis of high-quality and robust speech. 2) Our synthesized speech is comparable to the ground-truth speech in LibriSpeech and surpasses that in VCTK. The higher WER results in VCTK may stem from a noisy environment and the lack of ASR model fine-tuning in that dataset.

\begin{table*}[ht] \small
\centering
\caption{The robustness of NaturalSpeech 2 and other autoregressive/non-autoregressive models on 50 particularly hard sentences. We conduct an intelligibility test on these sentences and measure the number of word repeating, word skipping, and error sentences. Each kind of word error is counted at once per sentence.}
\begin{tabular}{ll|cc|cc}
\toprule
AR/NAR &Model & Repeats & Skips & Error Sentences & Error Rate \\
\midrule
\multirow{2}{*}{AR} & Tacotron~\cite{wang2017tacotron} & 4 & 11 & 12 & 24\%  \\
& Transformer TTS~\cite{li2019neural} & 7 & 15 & 17& 34\% \\
\midrule
\multirow{2}{*}{NAR} & FastSpeech~\cite{ren2019fastspeech} & 0 & 0 & 0 & 0\% \\
&  NaturalSpeech~\cite{tan2022naturalspeech} & 0 & 0 & 0 & 0\% \\ 
\midrule
\midrule
NAR & NaturalSpeech 2 & 0 & 0 & 0 & 0\% \\
\bottomrule
\end{tabular}
\label{table:robustness_evaluation}
\vspace{-5px}
\end{table*}

Autoregressive TTS models often suffer from alignment mismatch between phoneme and speech, resulting in severe word repeating and skipping. To further evaluate the robustness of the diffusion-based TTS model, we adopt the 50 particularly hard sentences in FastSpeech~\cite{ren2019fastspeech} to evaluate the robustness of the TTS systems. We can find that the non-autoregressive models such as FastSpeech~\cite{ren2019fastspeech}, NaturalSpeech~\cite{tan2022naturalspeech}, and also NaturalSpeech 2 are robust for the 50 hard cases, without any intelligibility issues. As a comparison, the autoregressive models such as Tacotron~\cite{wang2017tacotron}, Transformer TTS~\cite{li2019neural}, and VALL-E~\cite{wang2023neural} will have a high error rate on these hard sentences.  The comparison results are provided in Table \ref{table:robustness_evaluation}.

\subsection{Comparison with Other TTS Systems}

\begin{wraptable}{r}{0.38\textwidth} \small
\centering
\vspace{-10px}
\caption{SMOS and CMOS results between NaturalSpeech 2 and VALL-E.}
    \begin{tabular}{lcc}
    \toprule
    Setting & SMOS & CMOS \\
    \midrule
    GroundTruth & 4.09 & - \\
    \midrule
    VALL-E  &  3.53  & $-0.31$ \\
    NaturalSpeech 2 &  \textbf{3.83}  & $~~~\textbf{0.00}$ \\  
      \bottomrule
    \end{tabular}
\vspace{-10px}
\label{table:exp_valle_vs_ns_cmos}
\end{wraptable}
In this section, we compare NaturalSpeech 2 with the zero-shot TTS model VALL-E~\cite{wang2023neural}. We directly download  the first 16 utterances from VALL-E demo page\footnote{\url{https://valle-demo.github.io/}}, which consists of 8 samples from LibriSpeech and 8 samples from VCTK. We evaluate the CMOS and SMOS in Table \ref{table:exp_valle_vs_ns_cmos}.

From the results, we find that NaturalSpeech 2 outperforms VALL-E by 0.3 in SMOS and 0.31 in CMOS, respectively. The SMOS results show that \myname{} is significantly better in speaker similarity. The CMOS results demonstrate that the speech generated by NaturalSpeech 2 is much more natural and of higher quality.

\subsection{Ablation Study}
\label{sec:ablation_study}

\begin{table}[ht] \small
\centering
\caption{The ablation study of NaturalSpeech 2. The prosody similarity between the synthesized and prompt speech in terms of the difference in the mean (Mean), standard variation (Std), skewness (Skew), and kurtosis (Kurt) of pitch and duration. ``-" denotes the model can not converge.}
\begin{tabular}{lcccccccc}
\toprule
& \multicolumn{4}{c}{Pitch} & \multicolumn{4}{c}{Duration} \\
\cmidrule{2-5} \cmidrule{6-9}
& Mean$\downarrow$ & Std$\downarrow$ & Skew$\downarrow$ & Kurt$\downarrow$ &  Mean$\downarrow$ & Std$\downarrow$ & Skew$\downarrow$ & Kurt$\downarrow$ \\
\midrule
NaturalSpeech 2 & \textbf{10.11} & \textbf{6.18} & \textbf{0.50} & \textbf{1.01} & \textbf{0.65} & \textbf{0.70} & \textbf{0.60} & \textbf{2.99}\\
\midrule
w/o. diff prompt &  - & - & - & - & - & - & - & -  \\
\midrule
w/o. dur/pitch prompt & 21.69 &  19.38 & 0.63 & 1.29 & 0.77 & 0.72 & 0.70 & 3.70  \\
\midrule
w/o. CE loss  & 10.69 & 6.24 & 0.55 & 1.06 & 0.71 & 0.72 & 0.74 & 3.85 \\
\midrule
w/o. query attn & 10.78 & 6.29 & 0.62 & 1.37 & 0.67 & 0.71 & 0.69 & 3.59  \\
\bottomrule
\end{tabular}
\label{table:ablation_prosody_syn_prompt}
\end{table}

In this section, we perform ablation experiments. 1) To study the effect of the speech prompt, we remove the Q-K-V attention layers in the diffusion (abbr. w/o. diff prompt), and the duration and pitch predictors (abbr. w/o. dur/pitch prompt), respectively. 2) To study the effect of the cross-entropy 
(CE) loss $\mathcal{L}_{\rm ce-rvq}$ based on RVQ, we disable the CE loss by setting $\lambda_{ce-rvq}$ to $0$ (abbr. w/o. CE loss). 3) To study the effectiveness of two Q-K-V attention in speech prompting for diffusion in Section \ref{sec_method_incontext}, we remove the first attention that adopts $m$ randomly initialized query sequence to attend to the prompt hidden and directly use one Q-K-V attention to attend to the prompt hidden (abbr. w/o. query attn). We report the prosody similarity metric between synthesized and prompt speech in Table \ref{table:ablation_prosody_syn_prompt}. More ablation results between synthesized and ground-truth speech are included in Appendix~\ref{app:ablation_sty}.

We have the following observations: 1) Disabling speech prompt in the diffusion model significantly degrades prosody similarity (e.g., from $10.11$ to $21.69$ for the mean of the pitch or even can not converge), highlighting its importance for high-quality TTS synthesis. 2) Disabling cross-entropy loss worsens performance, as the residual vector quantizer's layer-wise cross entropy provides regularization for precise latent representations. 3) Disabling query attention strategy also degrades prosody similarity. In practice, we find that applying cross-attention to prompt hidden will leak details and thus mislead generation.

In addition, since the prompt length is an important hyper-parameter for zero-shot TTS, we would like to investigate the effect of the prompt length. We follow the setting of \textit{prosody similarity between synthesized and prompt speech} in Section \ref{sec:generation_similarity_exp}. Specifically, we vary the prompt length by $\sigma = \{3, 5, 10\}$ seconds and report the prosody similarity metrics of \myname{}. The results are shown in Table \ref{table:prosody_similarity_syn_prompt_length}. We observe that when the prompt is longer, the similarity between the generated speech and the prompt is higher for \myname{}. It shows that the longer prompt reveals more details of the prosody, which help the TTS model to generate more similar speech.

\begin{table}[ht]
\centering
\caption{The NaturalSpeech 2 prosody similarity between the synthesized and prompt speech with different lengths in terms of the difference in the mean (Mean), standard variation (Std), skewness (Skew), and kurtosis (Kurt) of pitch and duration.}
\begin{tabular}{lcccccccc}
\toprule
\multirow{3}{*}{\textbf{LibriSpeech}}& \multicolumn{4}{c}{Pitch} & \multicolumn{4}{c}{Duration} \\
\cmidrule{2-5} \cmidrule{6-9}
& Mean$\downarrow$ & Std$\downarrow$ & Skew$\downarrow$ & Kurt$\downarrow$ &  Mean$\downarrow$ & Std$\downarrow$ & Skew$\downarrow$ & Kurt$\downarrow$ \\
\midrule
3s & 10.11 & 6.18 & 0.50 & 1.01 & 0.65 & 0.70 & 0.60 & 2.99\\
\midrule
5s & 6.96 & 4.29 & 0.42 & 0.77 & 0.69 & 0.60 & 0.53 & 2.52  \\
\midrule
10s & 6.90 & 4.03 & 0.48 & 1.36 & 0.62 & 0.45 & 0.56 & 2.48  \\
\midrule
\midrule
\multirow{3}{*}{\textbf{VCTK}} & \multicolumn{4}{c}{Pitch} & \multicolumn{4}{c}{Duration} \\
\cmidrule{2-5} \cmidrule{6-9}
& Mean$\downarrow$ & Std$\downarrow$ & Skew$\downarrow$ & Kurt$\downarrow$ &  Mean$\downarrow$ & Std$\downarrow$ & Skew$\downarrow$ & Kurt$\downarrow$ \\
\midrule
3s & 13.29 & 6.41 & 0.68 & 1.27 & 0.79 & 0.76 & 0.76 & 2.65 \\
\midrule
5s & 14.46 &  5.47 & 0.63 & 1.23 & 0.62 & 0.67 & 0.74 & 3.40  \\
\midrule
10s & 10.28 & 4.31 & 0.41 & 0.87 & 0.71 & 0.62 & 0.76 & 3.48  \\
\bottomrule
\end{tabular}
\label{table:prosody_similarity_syn_prompt_length}
\end{table}

\subsection{Zero-Shot Singing Synthesis}
\label{sec_results_singing}
In this section, we explore NaturalSpeech 2 to synthesize singing voice in a zero-shot setting, either given a singing prompt or only a speech prompt. 

For singing data collection, we crawl a number of singing voices and their paired lyrics from the Web. For singing data preprocessing,  we utilize a speech processing model to remove the backing vocal and accompaniment in the song, and an ASR model to filter out samples with misalignments. The dataset is then constructed using the same process as speech data, ultimately containing around $30$ hours of singing data. The dataset is upsampled and mixed with speech data for singing experiments.

We use speech and singing data together to train NaturalSpeech 2 with a $5e-5$ learning rate. In inference, we set the diffusion steps to $1000$ for better performance. To synthesize a singing voice, we use the ground-truth pitch and duration from another singing voice, and use different singing prompts to generate singing voices with different singer timbres. Interestingly, we find that NaturalSpeech 2 can generate a novel singing voice using speech as the prompt. See the demo page\footnote{\url{ https://speechresearch.github.io/naturalspeech2}} for zero-shot singing synthesis with either singing or speech as the prompt.

\subsection{Extension to Voice Conversion and Speech Enhancement}
In this section, we extend \myname{} to another two speech synthesis tasks: 1) voice conversion and 2) speech enhancement. See the demo page\footnote{\url{ https://speechresearch.github.io/naturalspeech2}} for zero-shot voice conversion and speech enhancement examples.

\subsubsection{Voice Conversion}
Besides zero-shot text-to-speech and singing synthesis, \myname{} also supports zero-shot voice conversion, which aims to convert the source audio $z_{source}$ into the target audio $z_{target}$ using the voice of the prompt audio $z_{prompt}$. 
Technically, we first convert the source audio $z_{source}$ into an informative Gaussian noise $z_1$ using a \textit{source-aware diffusion process} and generate the target audio $z_{target}$ using a \textit{target-aware denoising process}, shown as follows.

\paragraph{Source-Aware Diffusion Process.} 
In voice conversion, it is helpful to provide some necessary information from source audio for target audio in order to ease the generation process. Thus, instead of directly diffusing the source audio with some Gaussian noise, we diffuse the source audio into a starting point that still maintains some information in the source audio. Specifically, inspired by the stochastic encoding process in Diffusion Autoencoder~\cite{preechakul2022diffusion}, we obtain the starting point $z_1$ from $z_{source}$ as follows:
\begin{equation}
\label{eq:define_z1}
\begin{split}
    z_{1} &= z_{0} + \int_{0}^{1} -\frac{1}{2} (z_t + \Sigma_t^{-1}(\rho(\hat s_{\theta}(z_t, t, c), t)-z_t)) \beta_t ~\mathrm{d}t,
    \end{split}
\end{equation}
where $\Sigma_t^{-1}(\rho(\hat s_{\theta}(z_t, t, c), t)-z_t)$ is the predicted score at $t$. We can think of this process as the reverse of ODE (Equation \ref{eq_reverse_ode}) in the denoising process. 

\paragraph{Target-Aware Denoising Process.} Different from the TTS which starts from the random Gaussian noise, the denoising process of voice conversion starts from the $z_1$ obtained from the source-aware diffusion process. 
We run the standard denoising process as in the TTS setting to obtain the final target audio $z_{target}$, conditioned on $c$ and the prompt audio $z_{prompt}$, where $c$ is obtained from the phoneme and the duration sequence of the source audio and the predicted pitch sequence.

As a consequence, we observe that \myname{} is capable of producing speech that exhibits similar prosody to the source speech, while also replicating the timbre specified by the prompt.

\subsubsection{Speech Enhancement}
\myname{} can be extended to speech enhancement, which is similar to the extension of voice conversion. 
In this setting, we assume that we have the source audio $z_{source}^{\prime}$ which contains background noise ( $z^{\prime}$ denotes the audio with background noise), the prompt with background noise $z_{prompt}^{\prime}$ for \textit{the source-aware diffusion process}, and the prompt without background noise $z_{prompt}$ for \textit{target-aware denoising process}. Note that $z_{source}^{\prime}$ and $z_{prompt}^{\prime}$ have the same background noise.

To remove the background noise, firstly, we apply the \textit{source-aware diffusion process} by $z_{source}^{\prime}$ and $z_{prompt}^{\prime}$ and obtain the $z_1$ as in Equation~\ref{eq:define_z1}. The source audio's duration and pitch are utilized in this procedure. Secondly, we run the \textit{target-aware denoising process} to obtain the clean audio by $z_1$ and the clean prompt $z_{prompt}$. 
Specifically, we use the phoneme sequence, duration sequence, and pitch sequence of the source audio in this procedure. As a result, we find that \myname{} can  effectively eliminate background noise while simultaneously preserving crucial aspects such as prosody and timbre.

\section{Conclusion and Future Work}
In this paper, we develop NaturalSpeech 2, a TTS system that leverages a neural audio codec with continuous latent vectors and a latent diffusion model with non-autoregressive generation to enable natural and zero-shot text-to-speech synthesis. To facilitate in-context learning for zero-shot synthesis, we design a speech prompting mechanism in the duration/pitch predictor and the diffusion model. By scaling NaturalSpeech 2 to 400M model parameters, 44K hours of speech, and 5K speakers, it can synthesize speech with high expressiveness, robustness, fidelity, and strong zero-shot ability, outperforming previous TTS systems. For future work, we will explore efficient strategies such as consistency models~\cite{song2023consistency,ye2023comospeech} to speed up the diffusion model and explore large-scale speaking and singing voice training to enable more powerful mixed speaking/singing capability.

\textbf{Broader Impacts}: Since NaturalSpeech 2 could synthesize speech that maintains speaker identity, it may carry potential risks in misuse of the model, such as spoofing voice identification or impersonating a specific speaker. We conducted the experiments under the assumption that the user agree to be the target speaker in speech synthesis. If the model is generalized to unseen speakers in the real world, it should include a protocol to ensure that the speaker approves the use of their voice and a synthesized speech detection model.

\clearpage


\bibliographystyle{unsrt}
\bibliography{ref}

\appendix
\clearpage

\section{Model Details}
\label{append_model_details}
\begin{table}[th]
  \label{model-details}
  \centering
  \begin{tabular}{cccc }
    \toprule
    Module     & Configuration    & Value & \#Parameters \\
    \midrule
    \multirow{5}{*}{Audio Codec} &  Number of Residual VQ Blocks & 16  & \multirow{5}{*}{27M}\\
    & Codebook size & 1024 &\\
    & Codebook Dimension & 256& \\
    & Hop Size & 200 &\\
    & Similarity Metric & L2 &\\
    \midrule
    \multirow{6}{*}{Phoneme Encoder} &  Transformer Layer & 6 & \multirow{6}{*}{72M} \\
    & Attention Heads & 8 & \\
    & Hidden Size & 512 &\\
    & Conv1D Filter Size & 2048& \\ 
    & Conv1D Kernel Size & 9 &\\
        & Dropout & 0.2 & \\
    \midrule
    \multirow{6}{*}{Duration Predictor} &  Conv1D Layers & 30 & \multirow{6}{*}{34M}\\
    & Conv1D Kernel Size & 3 & \\
    &  Attention Layers & 10& \\
    & Attention Heads & 8 &\\
    & Hidden Size & 512 &\\
    & Dropout & 0.5 & \\
    \midrule
    \multirow{6}{*}{Pitch Predictor} &  Conv1D Layers & 30 & \multirow{6}{*}{50M}\\
    & Conv1D Kernel Size & 5 & \\
    &  Attention Layers & 10 &\\
    & Attention Heads & 8 &\\
    & Hidden Size & 512 &\\
    & Dropout & 0.5 &\\
    \midrule
    \multirow{6}{*}{Speech Prompt Encoder}& Transformer Layer  & 6 & \multirow{6}{*}{69M} \\
    & Attention Heads & 8 & \\
    & Hidden Size & 512 &\\
    & Conv1D Filter Size & 2048 &\\ 
    & Conv1D Kernel Size & 9 &\\
        & Dropout & 0.2 &\\
    \midrule
    \multirow{7}{*}{Diffusion Model}& WaveNet Layer  & 40  & \multirow{7}{*}{183M}\\
    & Attention Layers & 13 & \\
    & Attention Heads& 8&\\
    & Hidden Size  & 512 &\\
    & Query Tokens & 32 &\\
    & Query Token Dimension & 512 &\\
    & Dropout & 0.2 &\\
    \midrule
    \multicolumn{3}{c}{Total} & 435M \\    
    \bottomrule
  \end{tabular}
  \vspace{0.1cm}
  \caption{The detailed model configurations of NaturalSpeech 2.}
  \label{fig:model_config_details}
\end{table}

\section{The Details of WaveNet Architecture in the Diffusion Model}
\label{append_wavenet_architecture}
\begin{figure*}[h]
  \centering
  \includegraphics[page=4,width=1\columnwidth,trim=0cm 1.0cm 7.2cm 0cm,clip=true]{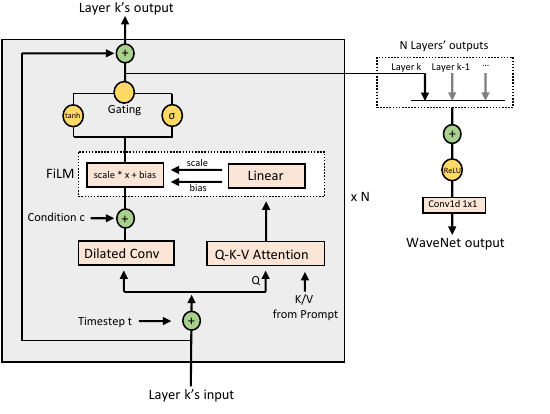}
  \caption{Overview of the WaveNet architecture in the diffusion model.}
  \label{fig_wavenet}
\end{figure*}

As shown in Figure~\ref{fig_wavenet}, the WaveNet consists of 40 blocks. Each block consists of 1) a dilated CNN with kernel size 3 and dilation 2, 2) a Q-K-V attention, and 3) a FiLM layer. In detail, we use Q-K-V attention to attend to the key/value obtained from the first Q-K-V attention module (from the speech prompt encoder) as shown in Figure~\ref{fig_incontext}. Then, we use the attention results to generate the scale and bias terms, which are used as the conditional information of the FiLM layer. Finally, we average the skip output results of each layer and calculate the final WaveNet output.

\section{The 50 Particularly Hard Sentences}
\label{appendix_50hard}
The 50 particularly hard sentences used in Section \ref{sec:robustness} are listed below:

\hspace*{\fill}
{
\setlength{\parskip}{0.2em}

\begin{hangparas}{1.6em}{1}
01. a \par
02. b \par
03. c \par
04. H \par
05. I \par
06. J \par
07. K \par
08. L \par
09. 22222222 hello 22222222 \par
10. S D S D Pass zero - zero Fail - zero to zero - zero - zero Cancelled - fifty nine to three - two - sixty four Total - fifty nine to three - two - \par
11. S D S D Pass - zero - zero - zero - zero Fail - zero - zero - zero - zero Cancelled - four hundred and sixteen - seventy six - \par
12. zero - one - one - two Cancelled - zero - zero - zero - zero Total - two hundred and eighty six - nineteen - seven - \par
13. forty one to five three hundred and eleven Fail - one - one to zero two Cancelled - zero - zero to zero zero Total - \par
14. zero zero one , MS03 - zero twenty five , MS03 - zero thirty two , MS03 - zero thirty nine , \par
15. 1b204928 zero zero zero zero zero zero zero zero zero zero zero zero zero zero one seven ole32 \par
16. zero zero zero zero zero zero zero zero two seven nine eight F three forty zero zero zero zero zero six four two eight zero one eight \par
17. c five eight zero three three nine a zero bf eight FALSE zero zero zero bba3add2 - c229 - 4cdb - \par
18. Calendaring agent failed with error code 0x80070005 while saving appointment . \par
19. Exit process - break ld - Load module - output ud - Unload module - ignore ser - System error - ignore ibp - Initial breakpoint - \par
20. Common DB connectors include the DB - nine , DB - fifteen , DB - nineteen , DB - twenty five , DB - thirty seven , and DB - fifty connectors . \par
21. To deliver interfaces that are significantly better suited  to create and process RFC eight twenty one , RFC eight twenty two , RFC nine seventy seven , and MIME content . \par
22. int1 , int2 , int3 , int4 , int5 , int6 , int7 , int8 , int9 , \par
23. seven \_ ctl00 ctl04 ctl01 ctl00 ctl00 \par
24. Http0XX , Http1XX , Http2XX , Http3XX , \par
25. config file must contain A , B , C , D , E , F , and G . \par
26. mondo - debug mondo - ship motif - debug motif - ship sts - debug sts - ship Comparing local files to checkpoint files ... \par
27. Rusbvts . dll Dsaccessbvts . dll Exchmembvt . dll Draino . dll Im trying to deploy a new topology , and I keep getting this error . \par
28. You can call me directly at four two five seven zero three seven three four four or my cell four two five four four four seven four seven four or send me a meeting request with all the appropriate information . \par
29. Failed zero point zero zero percent < one zero zero one zero zero zero zero Internal . Exchange . ContentFilter . BVT ContentFilter . BVT\_log . xml Error ! Filename not specified . \par
30. C colon backslash o one two f c p a r t y backslash d e v one two backslash oasys backslash legacy backslash web backslash HELP \par
31. src backslash mapi backslash t n e f d e c dot c dot o l d backslash backslash m o z a r t f one backslash e x five \par
32. copy backslash backslash j o h n f a n four backslash scratch backslash M i c r o s o f t dot S h a r e P o i n t dot \par
33. Take a look at h t t p colon slash slash w w w dot granite dot a b dot c a slash access slash email dot \par
34. backslash bin backslash premium backslash forms backslash r e g i o n a l o p t i o n s dot a s p x dot c s Raj , DJ , \par
35. Anuraag backslash backslash r a d u r five backslash d e b u g dot one eight zero nine underscore P R two h dot s t s contains \par
36. p l a t f o r m right bracket backslash left bracket f l a v o r right bracket backslash s e t u p dot e x e \par
37. backslash x eight six backslash Ship backslash zero backslash A d d r e s s B o o k dot C o n t a c t s A d d r e s \par
38. Mine is here backslash backslash g a b e h a l l hyphen m o t h r a backslash S v r underscore O f f i c e s v r \par
39. h t t p colon slash slash teams slash sites slash T A G slash default dot aspx As always , any feedback , comments , \par
40. two thousand and five h t t p colon slash slash news dot com dot com slash i slash n e slash f d slash two zero zero three slash f d \par
41. backslash i n t e r n a l dot e x c h a n g e dot m a n a g e m e n t dot s y s t e m m a n a g e \par
42. I think Rich's post highlights that we could have been more strategic about how the sum total of XBOX three hundred and sixtys were distributed . \par
43. 64X64 , 8K , one hundred and eighty four ASSEMBLY , DIGITAL VIDEO DISK DRIVE , INTERNAL , 8X , \par
44. So we are back to Extended MAPI and C++ because . Extended MAPI does not have a dual interface VB or VB .Net can read . \par
45. Thanks , Borge Trongmo Hi gurus , Could you help us E2K ASP guys with the following issue ? \par
46. Thanks J RGR Are you using the LDDM driver for this system or the in the build XDDM driver ? \par
47. Btw , you might remember me from our discussion about OWA automation and OWA readiness day a year ago . \par
48. empidtool . exe creates HKEY\_CURRENT\_USER \ Software \ Microsoft \ Office \ Common \ QMPersNum in the registry , queries AD , and the populate the registry with MS employment ID if available else an error code is logged . \par
49. Thursday, via a joint press release and Microsoft AI Blog, we will announce Microsoft's continued partnership with Shell leveraging cloud, AI, and collaboration technology to drive industry innovation and transformation. \par

50. Actress Fan Bingbing attends the screening of 'Ash Is Purest White (Jiang Hu Er Nv)' during the 71st annual Cannes Film Festival \par
\end{hangparas}
}

\section{Prosody Similarity with Ground Truth} 
\label{app:prosody_similarity_with_gt}
To further investigate the quality of prosody, we follow the generation quality evaluation of \textit{prosody similarity between synthesized and prompt speech} in Section~\ref{sec:generation_similarity_exp} and compare the generated speech with the ground-truth speech.
We use Pearson correlation and RMSE to measure the prosody matching between generated and ground-truth speech. The results are shown in Table \ref{table:prosody_similarity_syn_gt}. We observe that NaturalSpeech 2 outperforms the baseline YourTTS by a large margin, which shows that our NaturalSpeech 2 is much better in prosody similarity.

\begin{table}[ht]
\centering
\caption{The prosody similarity between the synthesized and ground-truth speech in terms of the correlation and RMSE on pitch and duration.}
\begin{tabular}{lcccc}
\toprule
\multirow{3}{*}{\textbf{LibriSpeech}} & \multicolumn{2}{c}{Pitch} & \multicolumn{2}{c}{Duration} \\
\cmidrule{2-3} \cmidrule{4-5}
& Correlation $\uparrow$ & RMSE $\downarrow$ & Correlation $\uparrow$ & RMSE $\downarrow$  \\
\midrule
YourTTS  
 & 0.77 & 51.78 & 0.52 & 3.24 \\
NaturalSpeech 2 & \textbf{0.81} & \textbf{47.72} & \textbf{0.65} & \textbf{2.72} \\
\midrule
\midrule
\multirow{3}{*}{\textbf{VCTK}} & \multicolumn{2}{c}{Pitch} & \multicolumn{2}{c}{Duration} \\
\cmidrule{2-3} \cmidrule{4-5}
& Correlation $\uparrow$ & RMSE $\downarrow$ & Correlation $\uparrow$ & RMSE $\downarrow$  \\
\midrule
YourTTS  
 & 0.82 & 42.63 & 0.55 & 2.55 \\
NaturalSpeech 2 & \textbf{0.87} & \textbf{39.83} & \textbf{0.64} & \textbf{2.50} \\
\bottomrule
\end{tabular}
\label{table:prosody_similarity_syn_gt}
\end{table}

\section{Ablation Study}
\label{app:ablation_sty}
In this section, we also compare the prosody similarity between audio generated by the ablation model and the ground-truth speech in Table \ref{table:ablation_prosody_syn_gt}. Similar to the results of comparing the audio generated by the ablation model and prompt speech, we also have the following observations. 1) The speech prompt is most important to the generation quality. 2) The cross-entropy and the query attention strategy are also helpful in high-quality speech synthesis.

\begin{table}[ht]
\centering
\caption{The ablation study of NaturalSpeech 2. The prosody similarity between the synthesized and ground-truth speech in terms of the correlation and RMSE on pitch and duration. ``-" denotes that the model can not converge.}
\begin{tabular}{lcccc}
\toprule
& \multicolumn{2}{c}{Pitch} & \multicolumn{2}{c}{Duration} \\
\cmidrule{2-5}
& Correlation $\uparrow$ & RMSE $\downarrow$ & Correlation $\uparrow$ & RMSE $\downarrow$  \\
\midrule
NaturalSpeech 2 & \textbf{0.81} & \textbf{47.72} & \textbf{0.65} & \textbf{2.72} \\
\midrule
w/o. diff prompt &  - & - & - & -   \\
\midrule
w/o. dur/pitch prompt & 0.80 &  55.00 & 0.59 & 2.76  \\
\midrule
w/o. CE loss  & 0.79 & 50.69 & 0.63 & 2.73  \\
\midrule
w/o. query attn & 0.79 & 50.65 & 0.63 & 2.73  \\
\bottomrule
\end{tabular}
\label{table:ablation_prosody_syn_gt}
\end{table}

\end{document}